\documentclass[twocolumn,twoside,slac_two]{revtex4}
\usepackage{graphicx}
\usepackage{fancyhdr}
\newcommand{\pks}{PKS~2155-304\ } 
\newcommand{\pksc}{PKS~2155-304} 
\newcommand{\Fermi}{\textit{Fermi} }
\newcommand{\Fermic}{\textit{Fermi}}
 \newcommand{\fvar}{\mathrm{F}_{\mathrm{var}}}

\begin{document}

\title{Monitoring the synchrotron and Compton emission of \pksc: one year of observations with \textit{RXTE} and \Fermi}

%

\author{D. Sanchez$^*$, B. Giebels$^*$ on behalf of the LAT collaboration}
\affiliation{$^*$Laboratoire Leprince-Ringuet, Ecole~Polytechnique, CNRS/IN2P3  }

\begin{abstract}
\pks is a well known GeV and TeV emitter and one of the brightest blazars in the \Fermi sky.
We present the results of one year of data taking with \Fermi, producing the longest lightcurve on this object. Together with a long duration X-ray monitoring program with \textit{RXTE}, these data give a better picture of the emission mechanisms for \pks on long time scales.

\end{abstract}

\maketitle

\thispagestyle{fancy}


\section{Introduction}
Blazars are active galactic nuclei (AGN) with a jet of relativistic plasma pointing with a few degrees of the observers line of sight \cite{bland78}. They emit electromagnetic radiation over 20 decades of energy. For these sources, the emission from relativistic electrons in the jet ( $1<\gamma_e <10^{6-7}$, where $\gamma_e$ is the electron Lorentz factor) dominates almost the full spectrum. Here we are interested in a subset of blazars which are detected at very high energies ($E>100$ GeV, hereafter the TeV domain), denoted the "TeV blazars". Up to now, 29 blazars have been detected in this domain\footnote{Online catalogs such as TeVCat \url{http://tevcat.uchicago.edu} presents updated view of the TeV sky.} and, for all these sources, the spectrum is well described by a power law of photon index greater than 2 in the TeV Domain

In leptonic emission models of TeV blazars (\cite{lepto1},\cite{lepto2}), X-rays are produced by synchrotron emission of electrons having the highest energies ($\gamma_e>10^4$) whereas $\gamma$-rays are the result of inverse-Compton (IC) scattering of seed photons which are yet to be determined. Correlations between the X-ray and TeV $\gamma$-ray fluxes have been reported in many blazars \cite{pks2155_chandra}, indicating that the X-ray and the TeV photons are produced by the same population of relativistic particles. Due to the low GeV flux, the link between the X-ray and the GeV emission (100 MeV-300 GeV) for those blazars was still poorly measured by EGRET. A better understanding of the relationship between those two populations of electrons will allow the study of the mechanisms, which produce the variability in the jet. Since February 2009, \pks is the subject of an observation campaign involving \textit{RXTE} and \Fermic, which aims is to characterize the electromagnetic emission simultaneously in the X-ray and GeV range on time scales of a week or more.

\section{Temporal analysis}
The \Fermi data were taken between MJD 54688 and MJD 55087 and were analyzed between 200 MeV and 300 GeV with Science Tools version V9R15P4 \cite{atw08}. Events reconstructed as photons with the highest probability (\texttt{DIFFUSE CLASS}), coming from a region of 10 degrees around the coordinates of \pks were selected. In order to measure the flux with a good accuracy, 58 time bins based on 7 days integrations were made. Each bin is determined, using a maximum likelihood method (\texttt{gtlike}).

\begin{figure}[t!]
   \begin{center}
     \includegraphics[width=85mm]{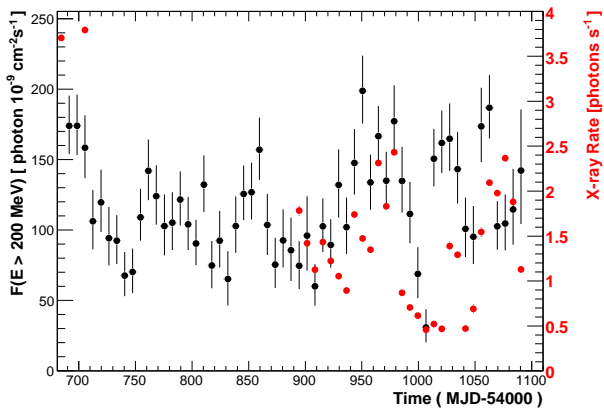}
  \end{center}
\label{LC}
\caption{7 days bin lightcurves of \pksc. In black, the \Fermi data and in red the \textit{RXTE} flux.}  
 \end{figure}

The \textit{RXTE} observations consist of 4 ks pointed observation every 3 days since MJD 54894, for a total of 64 observations. The \textit{RXTE} lightcurve was then rebinned to match the \Fermi one and allow a comparative analysis (Figure \ref{LC}). 

In order to quantify the source variability, we use the normalized excess variance \cite{fvar} defined by :

 \[
\fvar = \sqrt{\frac{S^2-\sigma^2_{err}}{\bar{f}^2}}\]

where $f$ is the observed flux with measurement error $\sigma_i^{err}$. $S^2$ is the variance of $f$ and $\sigma^2_{err}$ is the average of $\sigma_i^{err}$.

\begin{table}[t]
\begin{center}
\begin{tabular}{|c|c|c||c|c|}
\hline 
&2-6 {\rm keV}&6-12 {\rm keV}& 0.2-1 {\rm GeV} &1-300 {\rm GeV} \\ \hline
$\fvar$&$0.47 \pm 0.01$&$0.57 \pm 0.01$&$0.31 \pm 0.03 $&$0.25 \pm  0.04$ \\\hline
\end{tabular}
\caption{$\fvar$ as a function of the energy for 2 bands in the X-ray domain and GeV domain.}
\label{tablefvar}
\end{center}
\end{table}

The variability of \pks is energy dependent, with the X-rays ($\fvar^{\rm x}=0.44\pm 0.01$) showing greater variability than the $\gamma$-ray emission ($\fvar^\gamma=0.24 \pm 0.03$). This difference is however subject to bias since the time binning and observation methods are different. To better study the energy dependence of $\fvar$, lightcurves between 2-6 keV and 6-12 keV in X-rays and $200\,{\rm MeV}-1\,{\rm GeV}$ and $1-300\,{\rm GeV}$ for \Fermi were derived. The results of a variability analysis on these datasets are presented on Table \ref{tablefvar}.

In the X-ray domain, the 6-12 keV band is notably more variable than the 2-6 keV, which has been seen in other studies \cite{bib:421},\cite{Maraschi}. In the GeV domain, the values of $\fvar$ in the energy bands $200\,{\rm MeV}-1\,{\rm GeV}$ and $1-300\,{\rm GeV}$ are compatible with a constant, but the measurement error is large and more data are needed to better constrain a possible dependence of the variability with energy.

\begin{figure}[t!]
   \begin{center}
     \includegraphics[width=85mm]{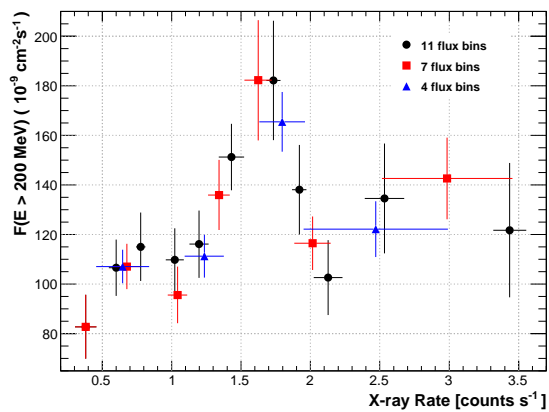}
   \end{center}
\label{FB}
\caption{\Fermi flux as a function of the X-ray flux. The \Fermi errors bars are statistical and the \textit{RXTE} one are equal to the variance of the time series.}
 \end{figure}

A correlation between the X-ray and GeV emission from \pks was previously observed during the joint H.E.S.S.-\Fermi multiwavelength campaign in August 2008 \cite{pks2155_2009} (Section \ref{secCamp}). During the current campaign, the two lightcurves also appear to be correlated, the Pearson correlation factor being $\rho = 0.39 \pm 0.08$. With the increased statistics available, a more detailed study was done in order to better understand the link between the two energy bands. Intervals corresponding to the times where the X-ray flux was between certain thresholds was defined. The \Fermi tools allow a standard analysis to be performed on these disjoint time segment. As an output, the \Fermi flux (and the statistical error) averaged on all those intervals is obtained. The analysis was then done for 4, 7 and 11 bins in flux, the results are shown in Figure \ref{FB}.

The three binnings give compatible results. The X-ray and $\gamma$-ray fluxes appear to be correlated when the X-ray flux is low, but not when the X-ray flux is higher. The transition appends at almost 1.7 counts by second in the PCA. These results confirm a X-ray-GeV correlation for \pks but show a non-trivial flux dependence. Nevertheless, during the campaign, the X-ray flux remained at a relative low level not allowing a confirmation of correlations during flaring episodes as was observed between the X-ray emission and the TeV $\gamma$-ray (\cite{pks2155_chandra}).


\section{Spectral energy distibution}
\begin{figure}[t!]
   \begin{center}
     \includegraphics[width=85mm]{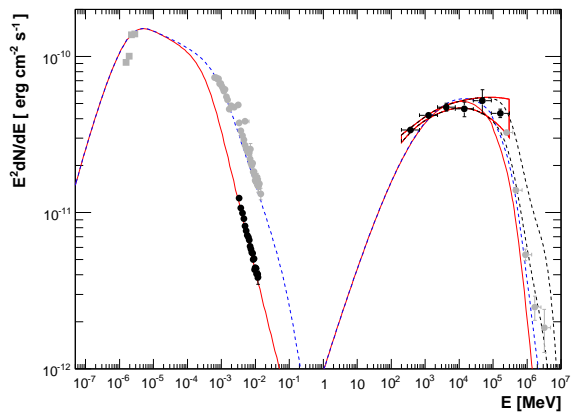}
  \end{center}
\label{SED}

\caption{Time-averaged SED. \textit{RXTE} and \Fermi data points are in black, Grey points are archival data from the MWL campaign in August 2008. In blue, the SSC model form \cite{pks2155_2009} and in red the same model adjusted for the present data.}
 \end{figure}

The time averaged \Fermi spectrum on the entire observation period is compatible with a log parabola function defined by :

\[F(E) = N_0 (E/E_0)^{-(\alpha +\beta~log_{10}(E/E_0))}\] 

where $E_0=300 {\rm MeV}$, $\alpha = 1.79\pm0.05$ and $\beta = 0.024\pm0.011$, for a total flux $F(>200\, {\rm MeV}) =(11.1 \pm 0.4)\times 10^{-8} \mathrm{ph\,cm}^{-2}\,\mathrm{s}^{-1}$, consistent with \cite{pks2155_2009}. A simple power law spectrum is ruled out at the 98 $\%$ confidence level. In the $\nu F\nu$ representation, the spectrum of \pks reaches a maximum around 20 GeV. The 1$\sigma$ error contour, shown in figure \ref{SED}, was extrapolated up to 10 TeV and corrected for absorption due to the extragalactic light (EBL) using the Franceschini model \cite{ebl}. This extrapolation is in good agreement wit the contemporaneous H.E.S.S data taken in August 2008 during a low state.
 
With a redshift of $z=0.117$, the effect of EBL on the spectrum of \pks is still negligible (absorption of 6 \%) at approximately 300 GeV, the maximum energy used in the fit, so the position of the peak likely to be intrinsic to the source. Two origins on the peak position could be proposed. The first possible effect is the IC emission mechanism which produce the $\gamma$ photons. When, in the rest frame of the comptonizing electron, the incident target photons have a very high energy, the relativistic Klein-Nishina (KN) effect leads to a decrease of the cross section, not allowing to the most energetic electrons to effectively upscatter photons. A second possibility is a diminution of the number of electrons above some cutoff energy. This threshold would be directly linked to the acceleration mechanism.

Between 2 an 10 keV, the spectrum is well described by a simple power law with a spectral index $\Gamma_X = 2.92\pm 0.01$ and a total flux  ${\rm F}_{2-10\,{\rm keV}} =2.6\times 10^{-11}\,\mathrm{erg\,cm}^{-2}\,\mathrm{s}^{-1}$. This flux is 2 times lower than the flux measured in August 2008.

The time averaged SED was adjusted with a one zone synchroton self-Compton (SSC) model similar to \cite{pks2155_2009}. The electron density is described by three powers law of indices $p_0,~p_1,~p_2$ between $\gamma=1,\gamma_1,\gamma_2$ respectively, and the same parameter values (B, R, $\delta$, etc...) were used, except $\gamma_2$ which is divided by a factor 1.5 and the slope  $p_2=4.9$ in order to describe the newer X-ray data. These changes to the parameters cause no notable modification to the IC emission leading to the conclusion that the most energetic electrons produced only a small amount of $\gamma$ photons. This favors the interpretation that the position of the IC peak position representation is due to KN effects and not to a cutoff at high energy in the electron spectrum.
\section{The first H.E.S.S-\Fermi multiwavelenght campaign : surprising results.}
\label{secCamp}
\begin{figure}[t!]
   \begin{center}
     \includegraphics[width=85mm]{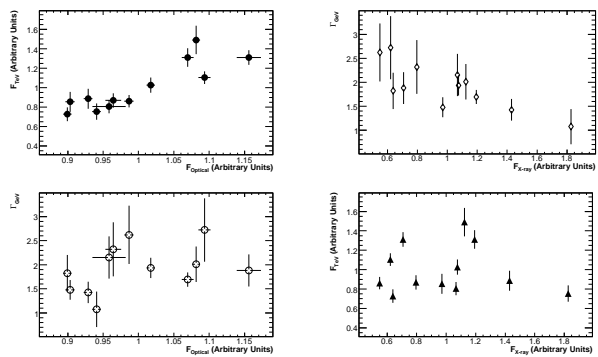}
   \end{center}
\label{MwL}
  \caption{Correlation patterns form the 2008 MWL campaign.}
 \end{figure}

In August 2008, \pks was the target of a MWL campaign. During 11 days, the electromagnetic spectrum has been covered by 5 instruments operating in 4 energy ranges : from optical with ATOM, through X-ray with \textit{Swift} and \textit{RXTE}, the GeV emission with \Fermi up to the TeV radiation with the H.E.S.S experiment. For the first time, the sensitivity of each instrument allowed to make lightcurves with 1-day bins and then study the correlation  between all the different wavebands. Figure \ref{MwL} gives a summary of the results of this campaign. Surprisingly, a correlation between the optical and TeV $\gamma$-ray fluxes and between the X-ray and GeV $\gamma$-ray fluxes have been, for the first time, observed. This correlation pattern is not simply explained with a one zone SSC model.

\vskip -.1cm

\section{Conclusions}
We show first results of an observation campaign of \pks with \textit{RXTE} and \Fermic. The lightcurves, extending over month, are correlated. Such results were already reported in August 2008 but, the X-ray flux, since February 2009, is two time lower. This fact as well as a more detailed study, realized with 7 months of data, show that the strength of the correlation is dependent of the X-ray flux. Nevertheless, more data are now accumulated and will allow to better understand the link between the X-ray photons produced by synchrotron emission and the soft $\gamma$-ray resulting of IC scattering. 

The \Fermi spectrum averaged on more than one year, has also allowed the shape of the peak at high energy to be constrained and the results  are in good agreement with the H.E.S.S measurement of the TeV flux of \pks in a low state. The picture that emerges from these campaigns is against the common idea that the most energetic electrons produce both the X-ray and TeV photons but suggest that the KN effect play a major role in the emission mechanism.

\section{Acknowledgements}
The \textit{Fermi} LAT Collaboration acknowledges generous ongoing support
from a number of agencies and institutes that have supported both the
development and the operation of the LAT as well as scientific data analysis.
These include the National Aeronautics and Space Administration and the
Department of Energy in the United States, the Commissariat \`a l'Energie Atomique
and the Centre National de la Recherche Scientifique / Institut National de Physique
Nucl\'eaire et de Physique des Particules in France, the Agenzia Spaziale Italiana
and the Istituto Nazionale di Fisica Nucleare in Italy, the Ministry of Education,
Culture, Sports, Science and Technology (MEXT), High Energy Accelerator Research
Organization (KEK) and Japan Aerospace Exploration Agency (JAXA) in Japan, and
the K.~A.~Wallenberg Foundation, the Swedish Research Council and the
Swedish National Space Board in Sweden.

Additional support for science analysis during the operations phase is gratefully
acknowledged from the Istituto Nazionale di Astrofisica in Italy and the Centre National d'\'Etudes Spatiales in France.

\end{document}